\documentclass[3p]{elsarticle}
\RequirePackage{amssymb}
\RequirePackage{longtable}

\makeatletter
\@addtoreset{equation}{section}
\makeatother

\newcounter{appnum}
\def\theappnum{\Alph{appnum}}
\makeatletter
\def\makeappendix#1{%
\stepcounter{appnum}
\addcontentsline{toc}{section}{\appendixname\space\theappnum #1}
\setcounter{equation}{0}
\section*{\appendixname\ \theappnum #1\@mkboth
 {\appendixname\ \theappnum}{\appendixname\ \theappnum}}
\let\thesection\theappnum}
\def\appendixname{Appendix}
\newcommand{\ccc}[1]{} 
\makeatother
\begin{document}


\title{Contemporary models of elastic nucleon scattering and their predictions for LHC}

\author[cern,fzu]{Jan Ka\v{s}par}
\ead{jan.kaspar@cern.ch}

\author[fzu]{Vojt\v{e}ch Kundr\'{a}t\corref{cor}}
\ead{kundrat@fzu.cz}

\author[fzu]{Milo\v{s} Lokaj\'{\i}\v{c}ek}
\ead{lokaj@fzu.cz}

\address[cern]{CERN, Geneva} 
\address[fzu]{Institute of Physics of the AS CR, v.v.i., 182 21 Prague 8, Czech Republic}

\cortext[cor]{Corresponding author}

\begin{abstract}
The analyses of elastic collisions of charged nucleons 
have been based standardly on West and Yennie formula. 
However, this approach has been shown recently to be 
inadequate from experimental as well as theoretical points 
of view. The eikonal model seems to be more pertinent as 
it enables to determine physical characteristics in impact 
parameter space. The contemporary phenomenological models  
cannot give, of course, any definite answer as the elastic 
collisions may be interpreted differently, as central or 
peripheral processes. Nevertheless, the predictions for 
the planned LHC energy have been given on their basis
and the possibility of exact determination of luminosity 
has been considered. 
\end{abstract}

\maketitle 

\section{Introduction}
\label{sec1}
The measurements of elastic scattering of charged 
nucleons at present high energies \cite{cart}-\cite{augi} 
have attained ample statistics enabling to perform very 
precise analyses of data measured in a broad interval 
of the four momentum transfer squared $t$ . The region 
of $t$'s where the differential cross section 
$d \sigma \over dt$ can be determined covers not 
only the region where nearly the pure 
hadron (nuclear) scattering is dominant, i.e.,
$\left|t\right| \gtrsim {10^{-2}}$ GeV$^2$, but 
also the region where the Coulomb scattering plays 
an important role, i.e., $|t| \lesssim 10^{-2}$ GeV$^2$ 
(the latter region being sometimes subdivided into 
Coulomb and interference parts). The complete scattering 
amplitude $F^{C+N}(s,t)$, fulfilling (in the normalization 
used by us) the relation 
\begin{equation}
{{d \sigma (s,t)} \over {dt}}= {\pi\over {sp^2}}|F^{C+N}(s,t)|^2
\label{ds1}
\end{equation}
has been commonly decomposed according to Bethe 
\cite{beth} into the sum of the Coulomb scattering
amplitude $F^{C}(s,t)$ known from QED and the hadronic 
amplitude $F^{N}(s,t)$ bound mutually by a relative 
phase $\alpha\Psi(s,t)$:
\begin{equation}
F^{C+N}(s,t) =
e^{i\alpha \Psi(s,t)} F^{C}(s,t)+ F^{N}(s,t);
\label{to1}
\end{equation}
$s$ is the square of the center of mass energy, 
$p$ is the momentum of an incident nucleon in the same 
system and $\alpha=1/137.036$ is the fine structure 
constant. The influence of spins of all particles involved 
in the elastic scattering has been neglected at the highest 
energies. 

The complete elastic scattering amplitude $F^{C+N}(s,t)$
used in the past has been established by West and Yennie 
\cite{west} and equals in the first approximation to
\begin{equation}
F^{C+N}(s,t) =
 \pm {\alpha s \over t} f_1(t)f_2(t)e^{i\alpha \Psi(s,t)}+
{\sigma_{tot}(s) \over {4\pi}} p\sqrt {s} 
\left ( \rho(s)+i\right ) e^{B(s)t/2}.
\label{wy1}
\end{equation}
The first term corresponds to the Coulomb scattering 
amplitude while the second term represents the elastic hadronic
amplitude. The upper (lower) sign corresponds to the scattering
of particles with the same (opposite) charges. The two form 
factors $f_1(t)$ and $f_2(t)$ in Eq.~(\ref{wy1}) describe
the electromagnetic structure of each nucleon (commonly in 
a dipole form) as
\begin{equation}
f_{j}(t) = \left(1+{|t|\over 0.71\,\rm GeV^2}\right)^{-2}.                  
\label{df1}
\end{equation}

Formula (\ref{wy1}) is valid provided the hadronic elastic 
amplitude (the second term on its right hand side) has a constant
diffractive slope $B$ together with constant quantity $\rho$
(the ratio of the real to imaginary parts of hadronic amplitude
in forward direction). Similarly as 
the total cross section $\sigma_{tot}$ they can depend only on
the energy. The relative phase $\alpha \Psi(s,t)$ in Eq.~(\ref{wy1}) 
has been shown by West and Yennie \cite{west} and independently
by Locher \cite{loch} to be
\begin{equation}
 \alpha \Psi(s,t) = \mp \alpha (\ln (-B(s)t/2) + \gamma )
 \label{wy2}
\end{equation}
where $\gamma=0.577215$ is the Euler constant.

Formulas (\ref{wy1}) and (\ref{wy2}) have been used 
for fitting the experimental data of differential 
cross section for small $|t|$ values (in the Coulomb, 
interference and also in a small adjacent part of 
hadronic domain) and the three mentioned quantities 
$\sigma_{tot}$, $B$ and $\rho$ have been determined. At 
larger $|t|$ values (i.e., in the hadronic region) the 
influence of Coulomb scattering has been usually fully 
neglected and elastic scattering has been described with 
the help of phenomenological elastic hadronic amplitude 
$F^N(s,t)$ which usually has exhibited a much more 
complicated $t$ dependence in this hadronic region than 
in Eq.~(\ref{wy1}). The different regions of differential 
cross section have been described by {\it{two different 
formulas}} (moreover based on incompatible assumptions) 
which has been recognized as important deficiency.

In the following section (Sec. \ref{sec2}) 
we will show in more detail which 
assumptions the formulas (\ref{wy1})-(\ref{wy2})
are based on and what are the limits of their use
in analyses of contemporary experimental data. In 
Sec. \ref{sec3} we will then discuss the approach 
based on the eikonal model which not only 
removes the corresponding limitations but 
also which describes the common influence of 
both the Coulomb and hadronic interactions in 
the whole measured region of momentum 
transfers uniquely with only one formula 
for the complete elastic amplitude. The $t$ 
dependence of the elastic hadronic scattering amplitude
$F^{N}(s,t)$ derived from experimental data within the
eikonal model enables to determine some physical
characteristics in the framework of impact parameter
space.

The aim of the presented paper is then contained mainly in the 
next three sections. The eikonal model approach will be used
for analysis of four phenomenological models proposed for 
a description of the elastic $pp$ scattering at the nominal    
LHC energy of 14 TeV; the model predictions will be given
and discussed in Sec. \ref{sec4}. The problems connected with 
the estimation of luminosity on the basis of elastic nucleon 
scattering will be analyzed in Sec. \ref{sec5}. The calculated 
root-mean-square (RMS) values of total, elastic and inelastic 
impact parameters corresponding to individual analyzed models
will be given and discussed in Sec. \ref{sec6}. And the results 
obtained on the basis of our approach will be summarized 
and discussed in Sec. \ref{sec7}.

\section{The West and Yennie formula}
\label{sec2}

The original function $\Psi(s,t)$ entering into 
Eq.~(\ref{to1}) has been derived by West and 
Yennie \cite{west} within the framework of Feynman
diagram technique in the case of charged point-like 
particles and for $s \gg m^2$ ($m$ stands for nucleon 
mass) as
\begin{equation}
\Psi_{WY}(s,t)=-\ln{-s\over t}-\int_{-4p^2}^{0}{dt'\over |t'-t|}
\left[1-{F^N(s,t')\over F^N(s,t)}\right],
\label{wy0}
\end{equation}
which has been further simplified and 
Eqs.~(\ref{wy1})-(\ref{wy2}) have been obtained. However, 
simplified formulas (\ref{wy1})-(\ref{wy2}) could hardly be
considered as a fully adequate tool for analyzing elastic nucleon 
scattering data already in time when they were derived. The issue 
is that formula (\ref{wy0}) has contained the integration over all 
kinematically allowed values of $t$ while experimental data has only 
covered a limited interval of $t$. Some assumptions defining and 
limiting the $t$ dependence of the hadronic amplitude, i.e., its 
modulus and phase defined in our case as
\begin{equation}
F^{N}(s,t) = i |F^{N}(s,t)| e^ {-i \zeta^{N} (s,t)},
\label{nu1}
\end{equation}
has had to be accepted to enable the integration. As nothing 
was known about the diffractive structure in $d \sigma \over dt$
at that time, the two following crucial assumptions
have been accepted:
\begin{itemize}
\item{the $t$ dependence of the modulus of the elastic
hadronic amplitude is purely exponential for {\it {all}} 
kinematically allowed $t$ values,}
\item{both the real and imaginary parts of the elastic
hadronic amplitude exhibit the same $t$ dependence for 
{\it{all}} admitted  $t$ values.}
\end{itemize}
In addition to these crucial assumptions, 
some high energy approximations has been added
(see, e.g., Refs. \cite{west}-\cite{kunx}).
Then the complete scattering amplitude has been written 
in the simplified form \cite{west} (for details see Ref. 
\cite{kunx}). Even if the standard fits obtained in the 
Coulomb and interference domains may seem to be good 
one cannot be sure about the actual meaning of fitted 
parameters since the data for higher $|t|$ values have 
not been taken into account quite correctly. 

In some papers (see, e.g., Refs. \cite{pump,haim}) the 
complete scattering amplitude $F^{C+N}(s,t)$ has been, 
therefore, described with the help of Eq.~(\ref{wy1}) 
containing the standard West and Yennie phase (\ref{wy2}) 
and the elastic hadronic amplitude $F^N(s,t)$ (substituting 
the second term in Eq.~(\ref{wy1})) constructed on the basis 
of some phenomenological ideas deviating from the two 
assumptions under which Eqs.~(\ref{wy1}) and (\ref{wy2})
were derived. Such an approach may be regarded, however, 
as very approximate.  

It might seem that a correct way may be reverting back to
integral formula (\ref{wy0}) in combination with formulas 
(\ref{ds1})-(\ref{wy2}). However, that is not possible, 
either, if the phase $\Psi_{WY}(s,t)$ should be real. 
{\it{The relative phase factor $\Psi_{WY}(s,t)$ can be 
real only provided the phase of the hadronic amplitude 
$\zeta^N(s,t)$ is $t$ independent in the whole region of 
kinematically allowed $t$ values}} \cite{kuny}; i.e., the 
quantity $\rho(s,t)$ should be constant in the whole interval 
of $t$. The contemporary experimental data as well as 
the phenomenological models of high energy elastic 
nucleon scattering show, however, convincingly that 
the quantity $\rho$ cannot be $t$ independent. Therefore, 
one should conclude that also the integral formula
(\ref{wy0}) should be designated as inadequate for the
description of elastic hadronic scattering. It is 
necessary to give decisive preference to a new and
more suitable approach based on eikonal model. 
In the following we should like to demonstrate the 
possibilities and advantages of the eikonal model
which is more general and more appropriate than that 
of West and Yennie.
\section{Eikonal model approach and mean-squares of impact parameters}
\label{sec3}
The complete elastic scattering amplitude $F^{C+N}(s,t)$  
is related by Fourier-Bessel transformation to the complete 
elastic scattering eikonal $\delta^{C+N}(s,b)$ \cite{adac}
\begin{equation}
F^{C+N}(s,q^2=-t)= {s\over {4 \pi i}} \int\limits_{\Omega_b}d^2b
e^{i\vec{q}\vec{b}} \left[e^{2i\delta^{C+N}(s,b)}-1\right],
\label{eik1} 
\end{equation}
where $\Omega_b$ is the two-dimensional Euclidean space of 
the impact parameter $\vec b$.

When formula (\ref{eik1}) is to be applied at finite energies
some problems appear as the amplitude $F^{C+N}(s,t)$ is defined
in a finite region of $t$ only. Mathematically consistent 
use of Fourier-Bessel transformation requires, however, 
the existence of the reverse transformation. And it is 
necessary to take into account the values of elastic 
amplitude from unphysical region where the elastic 
hadronic amplitude is not defined; for details see Refs.
\cite{adac}). This issue has been resolved in a unique way
by Islam \cite{isla1,isla2} by analytically continuing
the elastic hadronic amplitude $F^N(s,t)$ from the 
physical to the unphysical region of $t$ ; see also 
Ref. \cite{kunkas}.

The individual eikonals may be defined as integrals 
of corresponding potentials \cite{isla3}; and due to 
their additivity also the complete elastic eikonal 
$\delta^{C+N}(s,b)$ may be expressed as the sum of 
both the Coulomb $\delta^C(s,b)$ and hadronic 
$\delta^N(s,b)$ eikonals at the same value of 
impact parameter $b$ \cite{fran}:
\begin{equation}
\delta^{C+N}(s,b) = \delta^C(s,b) + \delta^N(s,b).
\label{eik2}
\end{equation}
The complete elastic scattering amplitude can be then
written as \cite{fran}-\cite{kun3}
\begin{equation}
F^{C+N}(s,t) = F^C(s,t) + F^N(s,t) + {i\over {\pi s}}
\int\limits_{\Omega_{q'}} d^2q'
F^{C}(s,{q'^2})F^{N}(s,[\vec{q} - \vec{q'}]^2), 
\label{eik3}
\end{equation}
where $\Omega_q$ is the two-dimensional set of 
kinematically allowed vectors $\vec q$.

This equation containing the convolution integral 
differs substantially from Eq.~(\ref{to1}). In the
final form (valid at any $s$ and $t$) it may be 
written \cite{kun3} as  
\begin{equation}
F^{C+N}(s,t) = \pm {\alpha s\over t}f_1(t)f_2(t) +
F^{N}(s,t)\left [1\mp i\alpha G(s,t) \right ],  
\label{kl1}
\end{equation}
where
\begin{equation}
G(s,t) = \int\limits_{-4p^2}^0
dt'\left\{ \ln \left( {t'\over t} \right )
{d \over{dt'}}
\left[f_1(t')f_2(t')\right]
+ {1\over {2\pi}}\left [{F^{N}(s,t')\over F^{N}(s,t)}-1\right]
I(t,t')\right\},
\label{kl2}
\end{equation}
and
\begin{equation}
I(t,t')=\int\limits_0^{2\pi}d{\Phi^{\prime \prime}}
{f_1(t^{\prime \prime})f_2(t^{\prime \prime})\over t^{\prime \prime}}, \;\;
t^{\prime \prime}=t+t'+2\sqrt{tt'}\cos{\Phi}^{\prime \prime}.
\label{kl3}
\end{equation}
Instead of the $t$ independent quantities
$B$ and $\rho$, it is now necessary to consider
corresponding $t$ dependent quantities defined as
\begin{equation}
B(s,t)= {d\over {dt}}\left[\ln {d \sigma^{N}\over {dt}}\right] =
{2\over |F^{N}(s,t)|}{d\over {dt}}|F^{N}(s,t)|
\label{sl1}
\end{equation}
and
\begin{equation}
\rho (s,t) = {{\Re F^{N}(s,t)} \over {\Im F^{N}(s,t)}}.
\label{ro1}
\end{equation}
The total cross section derived with the help 
of the optical theorem is then
\begin{equation}
\sigma_{tot} (s) = {{4 \pi}\over {p \sqrt{s}}} \Im F^{N}(s,t=0).
\label{tot1}
\end{equation}

The form factors $f_1(t)$ and $f_2(t)$ reflect the 
electromagnetic structure of colliding nucleons and form 
a part of the Coulomb amplitude from the very beginning.
But instead of using the dipole form factor (\ref{df1}) 
as it has been done in Eq.~(\ref{wy1}) it has been suggested
to use more convenient formula from Ref. \cite{bork}:
\begin{equation}
f_{j}(t) = \sum_{k=1}^4 {g_k\over{w_k-t}},
\hspace*{0.5cm}j=1,2
\label{bo1}
\end{equation}
where the values of the parameters $g_k$ and $w_k$ 
are to be taken from the quoted paper.

As the Coulomb part in formula (\ref{kl1}) is known
the complete amplitude depends in principle on hadronic 
amplitude $F^N(s,t)$ only. Thus it can be used in two 
complementary ways:
\begin{itemize}
\item{one can test the predictions of different models of 
high-energy elastic hadronic scattering that provide hadronic
amplitudes $F^{N}(s,t)$. Then, with the help of formula (\ref{kl1}) 
one can calculate complete amplitudes $F^{C+N}(s,t)$ that can be
compared to experimental data by employing Eq.~(\ref{ds1}),}

\item{one may resolve phenomenological $t$ dependence of 
elastic hadronic amplitude $F^{N}(s,t)$ at a given  $s$  
(and for {\it{all measured  t  values}}), by fitting
experimental elastic differential cross section data
with the help of Eq.~(\ref{ds1}) and (\ref{kl1}). The
crucial point here is then a suitable parameterization 
of the hadronic amplitude $F^N(s,t)$.}

\end{itemize}

The eikonal approach brings the possibility of determining 
mean values of impact parameter for different kinds of 
scattering processes. These quantities characterize the 
ranges of forces responsible for the elastic, inelastic 
and total scattering. If the unitarity condition and 
the optical theorem are applied to the mean-squared 
values of impact parameter for different processes 
may be determined directly from the $t$ dependence 
of elastic hadronic amplitude $F^N(s,t)$.

The elastic mean-square can be determined by means of 
the formula (see Refs. \cite{kunkas}, \cite{hene}-\cite{kunp})
\begin{eqnarray}
\!\!\!\!\!\!\!\!\!<b^2(s)>_{el} \!  &=& \!
4 \; {{\int \limits_{t_{min}}^{0}\!\! dt \;|t| \;
\left({d \over{dt}} |F^{N}(s,t)|\right)^2}
\over { \int \limits_{t_{min}}^{0}\!\! dt \; |F^{N}(s,t)|^2}} +
4 \; {{\int \limits_{t_{min}}^{0}\!\! dt \;|t| \;
|F^{N}(s,t)|^2 \left( {d \over {dt}} \zeta^{N}(s,t)\right)^2} 
\over { \int \limits_{t_{min}}^{0} \!\!dt \; |F^{N}(s,t)|^2}} \equiv
\nonumber  \\ 
& \equiv &  <b^2(s)>_{mod} + <b^2(s)>_{ph},  
\label{rmse}
\end{eqnarray}
where the modulus of elastic hadronic amplitude itself
determines the first term and the phase (its derivative) 
influences the second term only; note that both terms are
positive.

The total mean-square can be determined with the 
help of the optical theorem by (see Ref. \cite{kunr})
\begin{equation}
{\langle }b^2(s){\rangle }_{tot}\;\; = 2 B(s,0); 
\label{rmst}
\end{equation}
the diffractive slope $B(s,t)$ being defined by
Eq.~(\ref{sl1}).

According to the unitarity equation the averaged 
inelastic mean-square is related to the total and 
elastic mean-squares as \cite{kunr}
\begin{equation}
{\langle } b^2(s){\rangle }_{inel}\;\; = \; 
{{\sigma_{tot}(s)} \over {\sigma_{inel}(s)}} {\langle } 
b^2(s){\rangle }_{tot} \; - \;
{{\sigma_{el}(s)} \over {\sigma_{inel}(s)}} {\langle } b^2(s){\rangle
}_{el}.
\label{rmsi}
\end{equation}

\section{Model predictions for $pp$ elastic scattering at 
the nominal LHC energy}
\label{sec4}
In connection with the TOTEM \cite{tot1,tot2} and the ATLAS 
ALFA \cite{atlx} experiments where elastic $pp$ scattering  
will be studied, the predictions of four models proposed 
by Islam et al. \cite{isla}, Petrov, Predazzi and 
Prokhudin \cite{petr}, Bourrely, Soffer and Wu \cite{bou1} 
and Block, Gregores, Halzen and Pancheri \cite{blo2} will 
be discussed. Two different alternatives for the model of 
Petrov et al. \cite{petr} with two pomerons (2P) and with 
three pomerons (3P) will be considered. The mentioned 
models contain some free parameters in the formulas 
describing their $s$ and $t$ dependences. Their values
can be found in the quoted papers. The predictions for 
the nominal energy of 14 TeV are shown in Fig. 
\ref{fig:kasp_sigma,narrow} (small $|t|$ region) 
and Fig. \ref{fig:kasp_sigma,large} (large $|t|$ range).

\begin{figure}[h!]
\centerline{
\begin{tabular}{cc}
\includegraphics{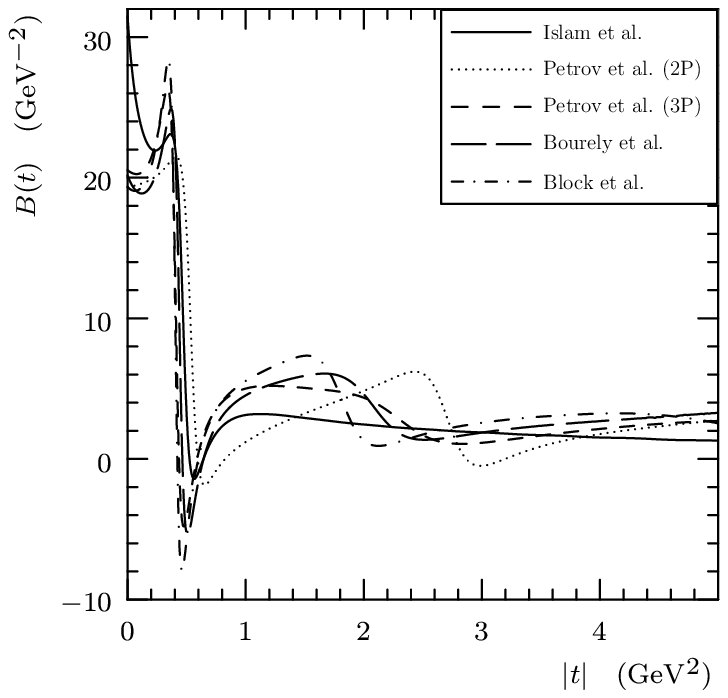} & 
\includegraphics{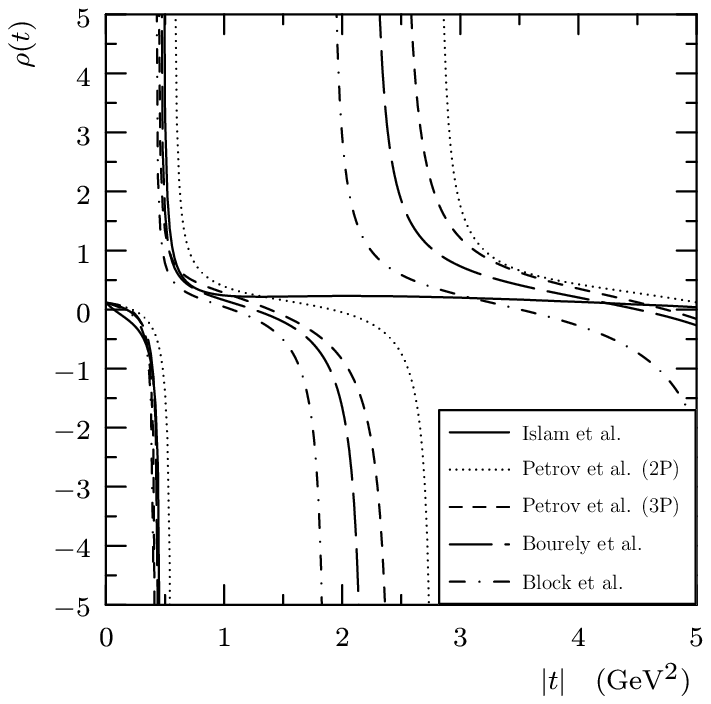}\\[-0.5cm]
\begin{minipage}[t]{0.45\textwidth}
\caption{\label{fig:kasp_slope}
The diffractive slope predictions for $pp$ scattering at 14 TeV
according to different models.} 
\end{minipage} & \begin{minipage}[t]{0.45\textwidth}
\caption{\label{fig:kasp_rho}
The $\rho (t)$ predictions for $pp$ scattering at 14 TeV
according to different models.} 
\end{minipage}\\
&\\[-0.5cm]
\end{tabular}}
\end{figure}

The total cross section $\sigma_{tot}(s)$, the 
diffractive slope $B(s,t)$ and the quantity 
$\rho(s,t)$ have been determined with the help 
of formulas (\ref{tot1}), (\ref{sl1}) and (\ref{ro1})  
for each model. The integrated elastic hadronic cross 
sections have been determined by integration of modified
Eq.~(\ref{ds1}) containing only $F^{N}(s,t)$.
The values of all these quantities are given in Table 1;
the corresponding graphs are shown in Figs. \ref{fig:kasp_slope} 
- \ref{fig:kasp_rho}. 
It is evident that the predictions of
divers models differ rather significantly;
the total cross section predictions range from 
95 mb to 110 mb. Another value of 101.5 mb 
following from the formula
\begin{equation}
\sigma_{tot}(s) = 21.70 \; \left({{s}\over{s_0}}\right)^{0.0808} 
                + \;\; 56.08 \; \left({{s}\over{s_0}}\right)^{-0.4525} \; mb,
                \hspace*{1cm} s_0 = 1 \; GeV^2
\label{dl1}
\end{equation}
has been given by Donnachie and Landshoff \cite{donn}
with the help of Regge pole model fit of $pp$
total cross sections performed at lower energies.
A higher value of $\sigma_{tot}$ has 
been established by COMPETE collaboration \cite{cude}
$\sigma_{tot} =  111.5\; \pm \; 1.2 ^ {\;+4.1} _{\;-2.1}$ 
mb which has been determined by extrapolation of the fitted 
lower energy data with the help of dispersion relations 
technique. Let us remark that there is no reliable theoretical 
prediction for this quantity: e.g., the latest prediction 
on the basis of QCD for this quantity has been $125 \pm 25$ mb
\cite{lan1}.
The predictions of ${d {\sigma} \over {dt}}$ values 
for higher values of $|t|$ are shown in Fig. 
\ref{fig:kasp_sigma,large}; they differ significantly 
for different models.
Let us point out especially the second diffractive 
dip being predicted by Bourrely, Soffer and Wu model 
\cite{bou1}. The predictions for the $t$ dependence 
of the diffractive slopes $B(t)$ are shown in Fig. 
\ref{fig:kasp_slope}. They differ significantly from the 
constant dependence required in the simplified West 
and Yennie formula (\ref{wy1}). Fig. \ref{fig:kasp_rho} 
displays the $t$ dependence of the quantity $\rho (t)$ 
that is not constant, either, as it would be required
by the second assumption needed for validity of formula 
(\ref{wy1}). Figs. \ref{fig:kasp_slope} and \ref{fig:kasp_rho} 
represent, 
\begin{center}
\begin{longtable}{ccccc}  
\hline  
     & & & &   \\
   model & $\sigma_{tot}$ & $\sigma_{el}$ & $B(0)$ & $\rho(0)$ \\ 
   & [mb] & [mb] & [GeV$^{-2}$] & \\
     & & & &   \\   
\hline \hline
     & & & &   \\  
   Islam  et al.     & 109.17 & 21.99 & 31.43 & 0.123 \\

Petrov et al. (2P)   &  94.97 & 23.94 & 19.34 & 0.097 \\

Petrov et al. (3P)   & 108.22 & 29.70 & 20.53 & 0.111 \\

Bourrely et al.      & 103.64 & 28.51 & 20.19 & 0.121 \\

Block et al.         & 106.74 & 30.66 & 19.35 & 0.114 \\
    & & & &   \\ 
\hline \\  
\caption{\label{tab:models1} The values of basic parameters predicted by
different models for $pp$ elastic scattering at energy of 14 TeV.}
\end{longtable} 
\end{center}
\vspace*{-1.0cm}
\begin{figure}[h!]
\centerline{
\begin{tabular}{cc}
\includegraphics{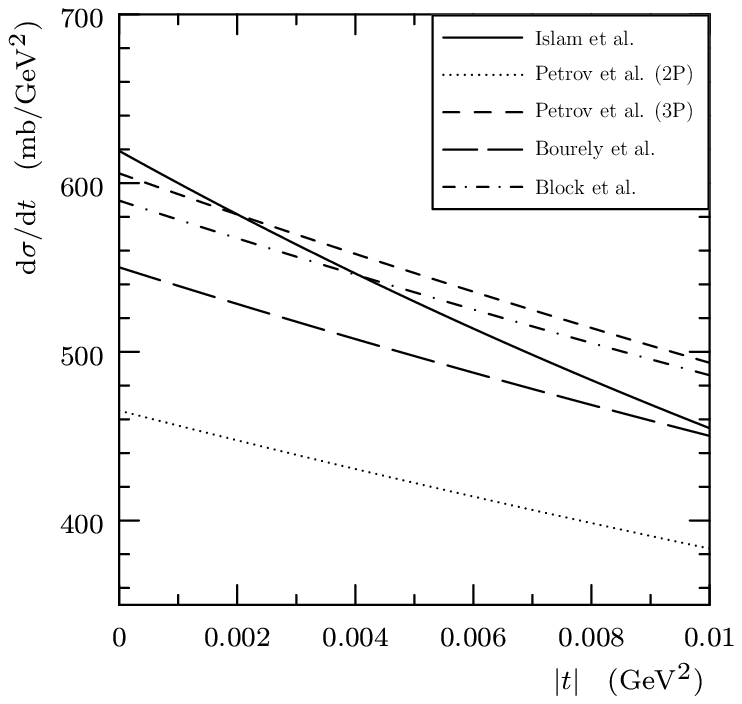} & 
\includegraphics{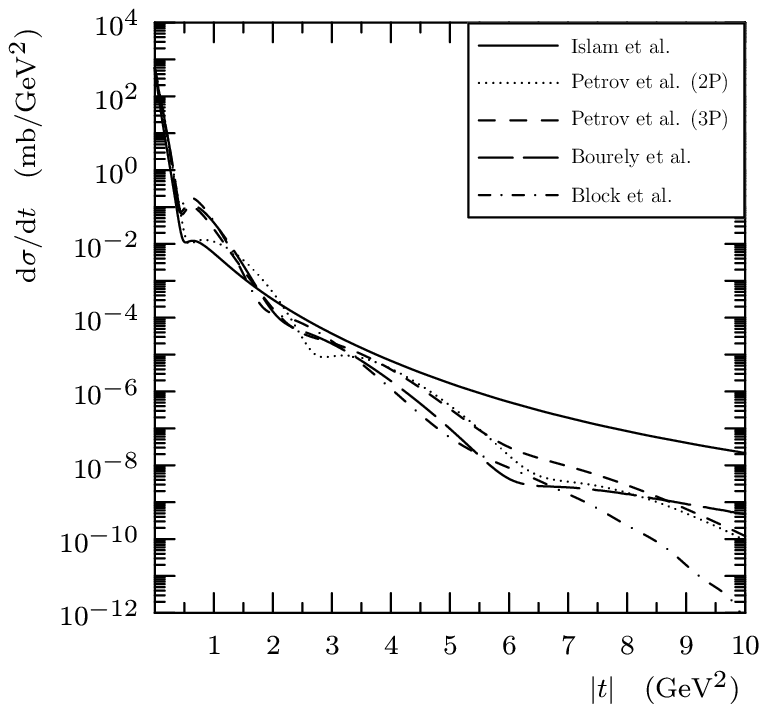}\\[-0.7cm]
\begin{minipage}[t]{0.45\textwidth}
\caption{\label{fig:kasp_sigma,narrow}
${{d \sigma} \over {dt}}$ predictions at low $|t|$
for $pp$ scattering at 14 TeV according to different models
(in very forward direction).} 
\end{minipage} & \begin{minipage}[t]{0.45\textwidth}
\caption{\label{fig:kasp_sigma,large}${{d \sigma} \over {dt}}$ 
predictions for $pp$ scattering at 14 TeV according to different models
(in a larger interval of $t$).} 
\end{minipage}\\
&\\[-0.5cm]
\end{tabular}}
\end{figure}
\begin{figure}[h!]
\centerline{
\begin{tabular}{cc}
\includegraphics{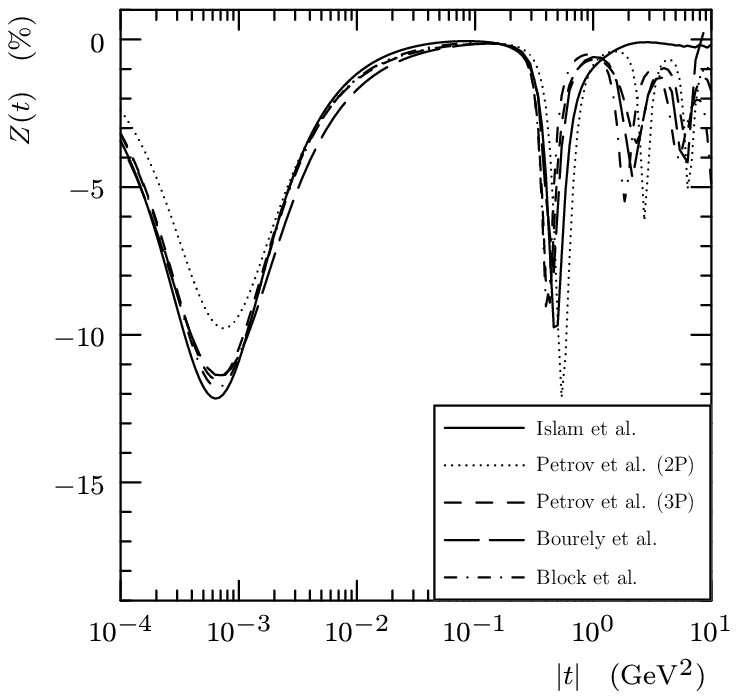} & 
\includegraphics{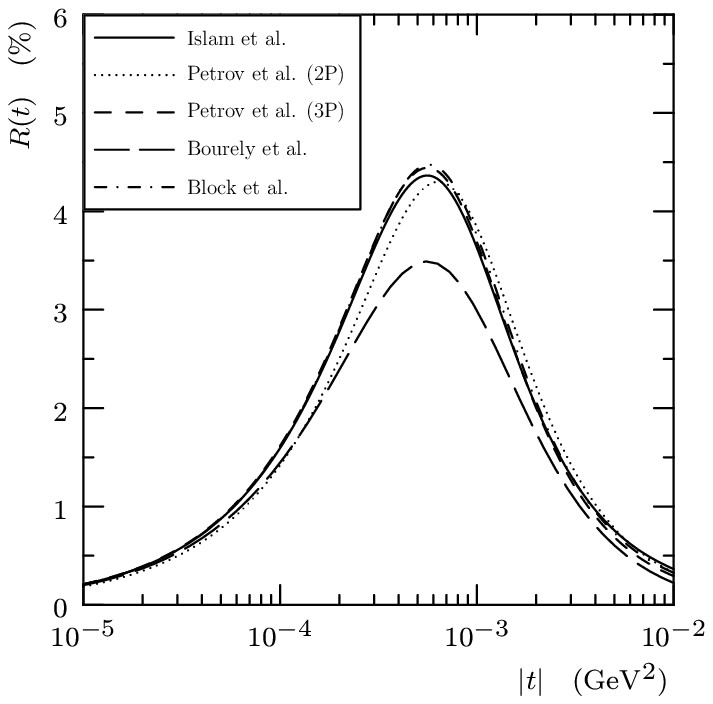}
\\[-0.7cm]
\begin{minipage}[t]{0.45\textwidth}
\caption{\label{fig:kasp_importance} The $t$ dependence of the ratio
of the interference to the hadronic contributions to the 
${{d \sigma}\over {dt}}$ for $pp$ elastic scattering at 14 TeV
according to different models.} 
\end{minipage} & \begin{minipage}[t]{0.45\textwidth}
\caption{\label{fig:kasp_R} The $R(t)$ quantity 
predictions for $pp$ scattering at 14 TeV
according to different models.} 
\end{minipage}\\
&\\[-0.5cm]
\end{tabular}}
\end{figure}
therefore, further support for the use of the 
eikonal formula for the complete elastic scattering 
amplitude (\ref{kl1}).
Fig. \ref{fig:kasp_importance} shows then the
$t$ dependence of the ratio of interference to
hadronic contributions of the ${{d \sigma}\over {dt}}$
for all of the given models, i.e., of the quantity
\begin{equation}
Z(t) \; = \; {{|F^{C+N}(s,t)|^2 - |F^{C}(s,t)|^2 - |F^{N}(s,t)|^2 }
\over {|F^{N}(s,t)|^2}}.
\end{equation}
The graphs show clearly that the influence of the 
Coulomb scattering may hardly be fully neglected 
also at higher values of $|t|$. It is interesting 
that at least for small $|t|$ the given characteristics 
are very similar.

\section{Luminosity estimation on the basis of $pp$ elastic scattering 
at the LHC}
\label{sec5}

An accurate determination of the elastic amplitude is very
important in the case when the luminosity of the collider 
is to be calibrated on the basis of elastic nucleon 
scattering. The luminosity $\mathcal{L}$ relates the
experimental elastic differential counting rate
${{d N_{el}}\over {dt}}(s,t)$ to the complete elastic
amplitude $F^{C+N}(s,t)$ (see Eq.~(\ref{ds1}) and Refs.
\cite{blo1,blo3}) by 
\begin{equation}
{1 \over \mathcal{L}} {{d N_{el}}\over {dt}}(s,t)  = {\pi \over {s p^2}}
 |F^{C+N} (s,t)|^2.
\label{lu1}
\end{equation}
Eq.~(\ref{lu1}) is valid for any admissible value of $t$. 
The value $\mathcal{L}$ might be in principle calibrated by
measuring the counting rate in the region of the smallest
$|t|$ where the Coulomb amplitude is dominant. However, 
this region can hardly be reached at the nominal LHC
energy due to technical limitations. A procedure allowing 
to avoid these difficulties may be based on Eq.~(\ref{lu1}), 
when the elastic counting rate may be, in principle, measured 
at any $t$ which can be reached, and the complete elastic
scattering amplitude $F^{C+N}(s,t)$ may be determined 
with required accuracy at any $|t|$, too. However,
in this case it will be very important which formula 
for the complete elastic amplitude $F^{C+N}(s,t)$ will 
be used. 

We have studied the differences between the West and Yennie 
simplified formula (see Eqs.~(\ref{wy1}) and (\ref{wy2}))
and the eikonal model (Eqs.~(\ref{kl1})-(\ref{kl3})).
The differences can be well visualized by the quantity
\begin{equation}
R(t) =  {{|F^{C+N}_{eik}(s,t)|^2 -
|F^{C+N}_{WY}(s,t)|^2}
\over {|F^{C+N}_{eik}(s,t)|^2}}, 
\label{lu2}
\end{equation}
where $F^{C+N}_{eik}(s,t)$ is the complete elastic 
scattering eikonal model amplitude, while $F^{C+N}_{WY}(s,t)$ 
is the  West and Yennie one. The quantity $R(t)$ is plotted 
in Fig. (\ref{fig:kasp_R}) for several models.

The maximum deviations lie approximately at \cite{blo1}
\begin{equation}
|t_{int}| \approx {{8 \pi \alpha} \over {\sigma_{tot}}} 
\approx 0.00064 \;\; GeV^2,
\label{cn1}
\end{equation}
where the Coulomb and the hadronic effects are expected
to be practically equal. Let us emphasize that the 
differences between the physically consistent eikonal 
model and the West and Yennie formula may reach 
almost 5 $\%$. It means that the luminosity derived 
on the basis of elastic $pp$ scattering at the energy 
of 14 TeV might be burdened by a non-negligible systematic 
error, if determined only from a small $t$ region around
$t_{int}$.

\begin{table}[htb]
\begin{center}
\begin{tabular}{cccc}  
\hline 
     & & &     \\
  & $\sqrt{<b^2_{tot}>}$ & $\sqrt{<b^2_{el}>}$ & $\sqrt{<b^2_{inel}>}$ \\ 
 model   & [fm] & [fm] & [fm] \\
       
     & & &     \\  
\hline \hline
     & & &    \\  
   Islam et al.     & 1.552 & 1.048 & 1.659 \\

Petrov et al. (2P)  & 1.227 & 0.875 & 1.324 \\

Petrov et al. (3P)  & 1.263 & 0.901 & 1.375 \\

Bourrely et al.     & 1.249 & 0.876 & 1.399 \\

Block et al.        & 1.223 & 0.883 & 1.336 \\
    & & &    \\ 
\hline \\  
\end{tabular}
\caption{\label{tab:models2} The values of root-mean-squares predicted by
different models.}
\end{center}
\end{table} 

\section{Root-mean-squared values of impact parameters}
\label{sec6}

The impact parameter representation of elastic hadronic
amplitude $F^N(s,t)$ allows to establish different
root-mean-squared (RMS) values of impact parameters 
that represent in principle the ranges of hadronic
interactions. Their values calculated with the help 
of formulas (\ref{rmse})-(\ref{rmsi}) for each of 
the analyzed models and expected at LHC nominal 
energy are shown in Table 2. The values 
of elastic RMS are in all cases lower than 
the corresponding values of the inelastic ones. 
It means that the elastic $pp$ collisions would 
be much more central then the inelastic ones; 
similarly as in the case of $pp$ scattering at the ISR 
energies for all these models; this should 
be recognized as a puzzle, see Ref. \cite{giac}. 
It can be interpreted as a consequence of admitting 
only a weak (standard) $t$  dependence of elastic 
hadronic phase in all models. The given puzzle can 
be removed if the used elastic hadronic phase 
$\zeta^N(s,t)$ is allowed to have a more general 
shape of $t$ dependence (see Refs. \cite{kunt,kunz,kuny,
kunkas,kun3,kunj}). 

While the $t$ dependence of modulus 
$|F^N(s,t)|$ can be determined from the measured 
elastic hadronic differential cross section the $t$ 
dependence of phase remains rather arbitrary 
(as already mentioned). And it is possible to choose
significantly different phase dependences \cite{kunr}.

It is, however, almost generally assumed that the imaginary 
part of elastic hadronic amplitude is dominant in a broad 
region of $|t|$ around the forward direction; it is taken
as slowly decreasing with rising $|t|$ and vanishing at
the diffractive minimum.  
The real part is assumed to start at 
small value at $|t|=0$ and to decrease, too, having still 
non-zero value at the diffractive minimum. It means that 
the $t$ dependence of the phase $\zeta^N(s,t)$ is very 
weak and becomes significant only in the region of diffractive
minimum. However, {\it{the existence of diffractive minimum 
does not require zero value for its imaginary part at this 
point. It means only that the sum of both the squares of  
real and imaginary parts should be minimal at this point.}} 
The mentioned requirement of vanishing imaginary part 
represents much stronger and more limiting condition then 
the physics requires.

Regarding Eq.~(\ref{rmse}) it is evident that 
very different elastic RMS values may be obtained
according to the chosen $t$ dependence of the phase
$\zeta^N(s,t)$. One should distinguish between the so 
called central picture (the first term dominates) and 
peripheral picture (decisive contribution comes rom the second 
term when the phase increases quickly with rising $t$ and
reaches $\pi/2$ at $|t| \simeq 0.1$ GeV$^2$). The value
$<b^2>_{el}$ is lesser than $<b^2>_{inel}$ in the central 
case while $<b^2>_{el}$ is greater than $<b^2>_{inel}$ 
in the peripheral case. The proton in the central case 
has been regarded as relatively transparent object 
which still represents a puzzling question (see, e.g., 
Refs. \cite{miet} and \cite{giac}).
And more detailed models of elastic hadronic scattering 
giving the peripheral distribution of elastic hadronic 
scattering should be considered and proposed. Only in 
such a case one may avoid the situation when the elastic 
hadronic scattering at high energies is more central than 
the inelastic ones as it follows immediately from the 
Fourier-Bessel transformation of elastic hadronic amplitude.
Thus no a priori limitations of elastic hadronic amplitude 
should be introduced in the corresponding analysis of 
experimental data and different possibilities should be
analyzed. 

As to the profiles in the impact parameter space
the peripheral behavior seems to be slightly preferred
on the basis of analysis of $pp$ experimental data at 53 GeV 
and $\bar{p}p$ at 541 GeV (see \cite{kun3}). The peripheral 
picture is supported also by analysis of elastic scattering 
of $\alpha$ particles on various targets 
($^1H, \; ^2H, \; ^3He, \; ^4He$) \cite{fran1} performed 
with the help of Glauber model where the 'elementary' 
nucleon-nucleon elastic hadronic amplitude has exhibited
similar $t$ dependence of phase $\zeta^N(s,t)$ as in our
peripheral case \cite{kun3}.

\section{Conclusion}
\label{sec7}
In the past the analyses of high energy elastic 
nucleon scattering data in the region of very small 
$|t|$ were performed with the help of the simplified 
interference formula proposed by West and Yennie and
including the influence of both Coulomb and hadronic 
interactions. At higher values of momentum transfers
the influence of Coulomb scattering was neglected
and the elastic scattering of nucleons was described 
only with the help of a hadronic amplitude having dominant 
imaginary part in a broad region of $t$ and vanishing
only at the diffractive minimum.  And it is evident that 
such a description of elastic nucleon scattering with the 
help of two different formulas for the complete amplitude 
represents significant deficiency. 

A more general eikonal model has been proposed. It describes 
elastic charged nucleon collisions at high energies with only 
one formula for the complete elastic amplitude in the whole 
kinematical region of $t$. This model is adequate for any 
$t$ dependence of the elastic hadronic
amplitude and has been successfully used for the analysis of 
elastic $pp$ and $\bar{p}p$ scattering data at lower energies - 
see, e.g., Ref. \cite{kun3}.

The attention of this paper has been devoted also 
to the LHC experiments that will measure proton-proton
elastic scattering \cite{tot2,atlx}. Several phenomenological
model predictions for dynamical quantities of interest have 
been discussed. A certain problem may be seen, however, in 
the fact that practically all considered allow central
behavior only.

Attention has been devoted also to the problem of luminosity 
determination as the values of all other quantities are affected 
by its value. The model predictions indicate that a systematic 
difference up to 5 $\%$ might occur between the eikonal and the
West and Yennie formulas.

It is also necessary to call attention to the fact that the
contribution of the Coulomb scattering cannot be fully
neglected at rather high $|t|$ values, either. However, 
the main open question concerns the fact that the
experimental data of the differential cross section give
directly the $t$ dependence of the modulus, while the $t$ 
dependence of the phase is only little constrained and may 
depend on some other assumptions or degrees of freedom. Any 
analysis of experimental data should, therefore, always contain 
a statistical evaluation of two different alternatives: 
central and peripheral; peripheral behavior corresponding 
better to usual picture of collision processes. And the 
attention should be devoted to a construction of the model 
which would be able to represent a realistic picture of 
elastic hadronic scattering of charged nucleons.

\vskip\baselineskip
\noindent{\bf{Acknowledgment:}} Valuable discussions with Prof. 
Karsten Eggert and Dr Mario Deile are highly appreciated.

\vskip\baselineskip


\begin{thebibliography}{999}

\bibitem{cart}
M.K. Carter, P.D.B. Collins and M. Whalley,
Compilation of nucleon - nucleon and nucleon - antinucleon 
elastic scattering data, 
Rutheford Lab. preprint, RAL-86-002 (1986)

\bibitem{bern}
D. Bernard et al.:
Phys. Lett. B 198 (1987) 583; 
M. Bozzo et al.: Phys. Lett. B 147 (1984) 385; B 155 (1985) 197

\bibitem{augi}
C. Augier et al.: 
Phys. Lett. B 316 (1993) 498

\bibitem{beth}
H. Bethe,
Ann. Phys. 3 (1958) 190

\bibitem{west}
G.B. West and D.R. Yennie,
Phys. Rev. 172 (1968) 1413

\bibitem{loch}
M.P. Locher,
Nucl. Phys. B2 (1967) 525
  
\bibitem{kunt}
V. Kundr\'{a}t, M. Lokaj\'{i}\v{c}ek,
Phys. Lett. B 232 (1989) 263

\bibitem{kunz}
V. Kundr\'{a}t and M. Lokaj\'{i}\v{c}ek,
Modern. Phys. Lett. A 11, No. 28 (1996) 2241

\bibitem{kunx}
V. Kundr\'{a}t and M. Lokaj\'{i}\v{c}ek,
Phys. Lett. B 611 (2005) 102

\bibitem{pump}
J. Pumplin,
Phys. Lett. B 276 (1992) 517

\bibitem{haim}
D. Haim and U. Maor,
Phys. Lett. B 278 (1992) 469
  
\bibitem{kuny}
V. Kundr\'{a}t, M. Lokaj\'{i}\v{c}ek and I. Vrko\v{c},
Phys. Lett. B 656 (2007) 182
 
\bibitem{adac}
T. Adachi and T. Kotani,
Progr. Theor. Phys. Suppl., Extra Number (1965) 316;
 37-38 (1966) 297;
Progr. Theor. Phys.  35 (1966)  463; 
 35 (1966) 485;
 39 (1968) 430;
 39 (1968) 785

\bibitem{isla1}
M.M. Islam,
Lectures in theoretical Physics, 
ed. A. O. Barut and W. E. Brittin, Vol. 10B
(Gordon and Breach, 1968), p.97

\bibitem{isla2}
M. M. Islam,
Nucl. Phys. B104 (1976) 511

\bibitem{kunkas}
V. Kundr\'{a}t, M. Lokaj\'{\i}\v{c}ek and J. Ka\v{s}par,
"Impact parameter profiles of elastic scattering
in hadronic collisions"; to be published

\bibitem{isla3}
M. M. Islam, 
Nuovo Cimento 48A (1967) 251

\bibitem{fran}
V. Franco,
Phys. Rev. D7 (1973) 215

\bibitem{cahn}
R. Cahn,
Z. Phys. C 15 (1982) 253

\bibitem{kun3}
V. Kundr\'{a}t and M. Lokaj\'{\i}\v{c}ek,
Z. Phys. C 63 (1994) 619;
see also CERN-Th.6952/93 preprint

\bibitem{bork}
F. Borkowski et al.,
Nucl. Phys. B93 (1975) 461

\bibitem{hene}
F. S. Heney and J. Pumplin,
Nucl. Phys. B117 (1976) 235

\bibitem{kunj}
V. Kundr\'{a}t, M. Lokaj\'{\i}\v{c}ek Jr. and M. Lokaj\'{\i}\v{c}ek,
Czech. J. Phys. B31 (1981) 1334

\bibitem{kunr}
V. Kundr\'{a}t and M. Lokaj\'{\i}\v{c}ek,
Phys. Lett. B 544 (2002) 132

\bibitem{kunp}
V. Kundr\'{a}t, M. Lokaj\'{\i}\v{c}ek and D. Krupa,
"Nucleon high-energy profiles", in Proceedings of the
IXth Blois Workshop on Elastic and Diffractive Scattering,
Pruhonice near Prague, Czech Republic, June 9-15, 2001,
Edits. V. Kundr\'{a}t, P. Z\'{a}vada, ISBN 80-238-8243-0

\bibitem{tot1}
V. Berardi et al., 
TOTEM Technical Design Report,
CERN-LHCC-2004-002 (2008)

\bibitem{tot2}
G. Anelli et al., 
The TOTEM experiment at the CERN Large hadron Collider,
2008 JINST 3 S08007

\bibitem{atlx}
ATLAS-ALFA: 
Technical Design Report,
CERN-LHCC-2004-004 (2008) 

\bibitem{isla}
M.M. Islam, R.J. Luddy and A.V. Prokhudin, 
Phys. Lett. B 605 (2005) 115

\bibitem{petr}
V.A. Petrov, E. Predazzi and A.V. Prokhudin,
Eur. Phys. J. C28 (2003) 525

\bibitem{bou1} 
C. Bourrely, J. Soffer and T. T. Wu, 
Mod. Phys. Lett. A6 (1991) 2973;
Mod. Phys. Lett. A7 (1992) 457(E);
Phys. Lett. B315 (1993) 195
Eur. Phys. J.28 (2003) 97

\bibitem{blo2}
M. M. Block, E. M. Gregores, F. Halzen and G. Pancheri,
Phys. Rev. D60 (1999) 0504024

\bibitem{donn}
A. Donnachie and P.V. Landshoff,
Phys. Lett. B 296 (1992) 227

\bibitem{cude}
J.R. Cudell et al., 
Phys. Rev. Lett. 89 (2002) 201801

\bibitem{lan1}
P.V. Landshoff,
Total cross sections, arXiv:0709.0395 [hep-ph]
 
\bibitem{blo1}
M.M. Block and R.N. Cahn, 
Rev. Mod. Phys. 57 (1985) 563

\bibitem{blo3}
M. M. Block,
Phys. Rept. 436 (2006) 71

\bibitem{miet}
H.G. Miettinen,
in proceedings of the IX th Rencontre de Moriond,
Meribel les Allues, Vol.1 (ed. J. Tran Thanh Van),
Orsay (1974)

\bibitem{giac}
G. Giacomelli and M. Jacob,
Phys. Rep. 55 (1979) 1

\bibitem{fran1} 
V. Franco and Y. Yin, 
Phys. Rev. Lett. 55 (1985) 1059; Phys. Rev. C 34 (1986) 608

\end{thebibliography}
\end{document}